*Title*:

# Automaticity in Computation and Student Success in Introductory Physical Science Courses

*Authors*:

JudithAnn R. Hartman,* Eric A. Nelson[†]

*Affiliations*:

*Department of Chemistry, United States Naval Academy, Annapolis, MD 21042 USA

[†]Fairfax County (VA, USA) Public Schools (retired)

*Corresponding Author:*

 *Hartman@usna.edu

*Keywords:*

Physics Education Research

Chemical Education Research

Mathematics

Problem Solving

Cognitive Science

Working Memory

(Discussion of this paper is invited in the comment section at www.ChemReview.Net/blog at Post 14.)

*Title*:

# Automaticity in Computation and Student Success in Introductory Physical Science Courses

*Abstract*:

Between 1984 and 2011, the percentage of US bachelor's degrees awarded in physics declined by 25%, in chemistry declined by 33%, and overall in physical sciences and engineering fell 40%. Data suggest that these declines are correlated to a K-12 de-emphasis in most states of practicing computation skills in mathematics.  Analysis of K-12 "state standards" put into place between 1990 and 2010 find that most states directed teachers to de-emphasize both memorization and student practice in computational problem solving.  Available state test score data show a significant decline in student computation skills.  In recent international testing, scores for US 16-24 year olds in numeracy finished last among 22 tested nations in the OECD. Recent studies in cognitive science have found that so solve well-structured problems in the sciences, students must first memorize fundamental facts and procedures in mathematics and science until they can be recalled "with automaticity," then practice applying those skills in a variety of distinctive contexts.  Actions are suggested to improve US STEM graduation rates by aligning US math and science curricula with the recommendations of cognitive science.

# Introduction

The success rates of US students in science, technology, engineering, and mathematics (STEM) courses are a matter of national concern (Augustine et al. 2010). In a 2012 report, the US President's Council of Advisors on Science and Technology (PCAST) cited "the rising demand for STEM talent in an increasingly technological world," and "a need for approximately 1 million more STEM professionals than the US will produce at the current rate over the next decade." (Olson and Riordan, 2012)

Despite the high demand for college graduates in STEM fields, the percentages of US students majoring in the physical sciences and engineering has been declining. Between 1984 to 2011, as a percentage of all US bachelor's degrees awarded, degrees in physics declined by 25%, in chemistry fell 33 % and overall degrees in the physical sciences and engineering declined 40% (Table 1) (National Science Foundation 2013, 2014).

| Year | Total US Bachelor's Degrees Awarded | Physics Bachelor's Degrees | Physics Bachelor's Percent of Total | Chemistry Bachelor's Degrees | Chemistry Bachelor's Percent of Total | Chemistry, Physics, and Engineering Bachelor's Degrees | Chemistry, Physics, and Engineering Percent |
|------|------|------|------|------|------|------|------|
| 1984 | 986,345 | 3,921 | 0.40% | 10,912 | 1.11 % | 90,986 | 9.22 % |
| 1993 | 1,179,278 | 4,080 | 0.35% | 9,109 | 0.77 % | 75,894 | 6.44 % |
| 2002 | 1,308,970 | 3,641 | 0.28% | 9,439 | 0.72 % | 73,685 | 5.63 % |
| 2011 | 1,734,229 | 5,221 | 0.30% | 12,888 | 0.74 % | 96,208 | 5.55 % |

**Table 1: Bachelor's Degrees by Major, 1984-2011**

These declines have occurred despite efforts to increase the number of majors in STEM fields. As one example, at US federal agencies in 2012, over 250 programs spent $2.5 billion to support STEM education (Kramer 2012).

Many of these initiatives aim to motivate students to study in STEM fields, but available data indicate that student interest in physical science and engineering majors has not substantially declined. In annual national polling by the UCLA Higher Education Research Institute between 1975 and 2008, entering college students stating their intention to major in the physical sciences and engineering remained relatively steady: between 10%-13% each year (National Science Board 2012, 2002).

What data show, rather than low interest, is low student survival rates in STEM majors. According to the 2012 PCAST report, for students entering college in 2003, the percentages who had attained a STEM degree of any kind six years later was 42% for students intending to major in engineering and 41% for students intending to major in the physical sciences.

Performance in introductory math and science courses appears to be a major barrier to majors in STEM fields. A 2008 study at SUNY Albany found that from 2003 to 2006, of 1,800 students who started General Chemistry, 60% earned a D or F or had withdrawn by the end of the two semester course. Of 964 students who started General Physics, 29% completed the two semester course with a C or above. (Cavanaugh 2008).

A 2013 study by the National Center on Education and the Economy found that at the community college level, "the workhorse of our postsecondary education system…, a large proportion of our high school graduates are unable to succeed in their first year." The report cited deficits in "middle school math" as a major factor in failure rates.

In a 1995 study at a polytechnic university with average SAT math scores for entering students of over 600, Knight found that "a full 50%" of students entered calculus-based introductory physics "with no useful knowledge of vectors at all." Testing students at a large US state university at the conclusion of one semester of introductory college physics, Nguyen and Meltzer (2003) found that over 50% of students completing algebra-based physics and over 25% of students completing calculus-based physics could not add vectors in two dimensions.

At a large state university with a high school grade point average for entering students of 3.7, Stanich, Craig, and Keller (2012) reported that approximately 20% of students in general chemistry were unable to solve exponential notation calculations and 40% to 50% were unable to solve logarithmic calculations.

Background knowledge in math has been found to be a strong predictor of success in introductory physical science courses. In 2006, Tai, Ward, and Sadler found that SAT-verbal scores, parental education level, ethnicity, high school coverage of chemistry topics, and enrollment in high school Advanced Placement (AP) Chemistry were all correlated with grades in first semester introductory college chemistry, but the best predictor identified was high school enrollment in AP mathematics. "Fluency in algebra" was cited as a likely explanation for the correlation.

Several studies have found a correlation between general chemistry success and "computational fluency:" The ability to solve simple calculations, using memorized facts and algorithms without a calculator. Among several measures, Wagner, Sasser, and DiBiase (2002) found the most reliable predictor of success in first semester general chemistry was a test of math calculations with no calculators allowed. Additional gains in identifying students who would be "at-risk" in general chemistry were reported by Cooper and Pearson (2012) testing both math and chemistry fundamentals without a calculator.

Leopold and Edgar (2008) reported a significant correlation between student final grades in second semester general chemistry and their success solving numerically simplified algebra, exponential notation, and base 10 logarithms without a calculator. They "argued that the results indicate an inadequate degree of mathematics fluency for the majority of the students tested."

Yet by some measures, the math skills of US students have been improving. The National Assessment of Educational Progress (NAEP), termed the "nation's report card," is administered to a large sample of US students in two different formats. On the "main NAEP" given in the 4th and 8th grade, 8th grade math scores have risen steadily. In 2011, 73% of students were found to have at least a basic knowledge of eighth-grade mathematics, compared to 52 percent in 1990 (National Center for Education Statistics 2011).

Some science educators have suggested that because calculators and software can now perform many of the typical calculations in math, physics, and chemistry, the need for students to master these operations has diminished (Cooper and Klymkowsky 2013). In contrast, a US Presidential Commission in 2008 cited cognitive studies finding that unless students had previously thoroughly memorized fundamentals of arithmetic and algebra, they would likely be unable to master solving scientific calculations (Geary et al. 2008).

Because the recent measures of math achievement have been inconsistent and their importance has been a subject of debate, we sought additional data on math background and its current relevance for learning in the quantitative sciences. Questions regarding the achievement of students in the mathematics needed for the sciences have been raised in other nations (Howson et al. 1995, Stokke 2015). For this paper, however, we will limit our scope to consideration of data for the United States and the question of student preparation to solve the "well-structured end-of-chapter" problems typically found in US college texts in introductory physics and chemistry for science majors.

In this study, we asked three related questions about instruction for students intending to major in physics, chemistry, or engineering:

1. Is it important that students be able to solve fundamental scientific calculations without reliance on computers or calculators?

2. In the US, how much has student math background been a factor in low introductory course success rates and the decline in degrees in the physical sciences and engineering?

3. Can we identify actions that may be the most promising in improving success in quantitative science courses?

Since the 1970's, many US states have used standardized tests to assess K-12 math skills. We began by examining those results.

## State Test Data

In Iowa, over 300 independent K-12 school districts adopt textbooks without guidance from the state. Between 1978 and 2001, Iowa did not have "state standards" for math, but all local districts were required to measure 8th grade student achievement using the *Iowa Test of Basic Skills* (ITBS), a nationally normed test administered during this period in many states (Am. Educational Research Association 1999). The "full" ITBS administered in Iowa consisted of three math subtests: "Concepts and Estimation," "Problems and Data Interpretation," and "Computation." The computation subtest involved "arithmetic operations with whole numbers, fractions, and decimals performed without a calculator" (Riverside Publishing 2011). Results, reported as grade level equivalents, were compared each year to fixed 1965 national norms for both "total math composite" and subtest scores (Iowa Dept. of Education 2002).

In 2002, Loveless compared Iowa ITBS combined scores for the two non-computation subtests to scores in computation. Between 1978 and 1990, for over 30,000 8th graders tested each year, scores rose on both measures. After 1991, "non-computation" was relatively level, but "computation" scores declined every year, falling in 2001 to their lowest level in 23 years [Figure 1].

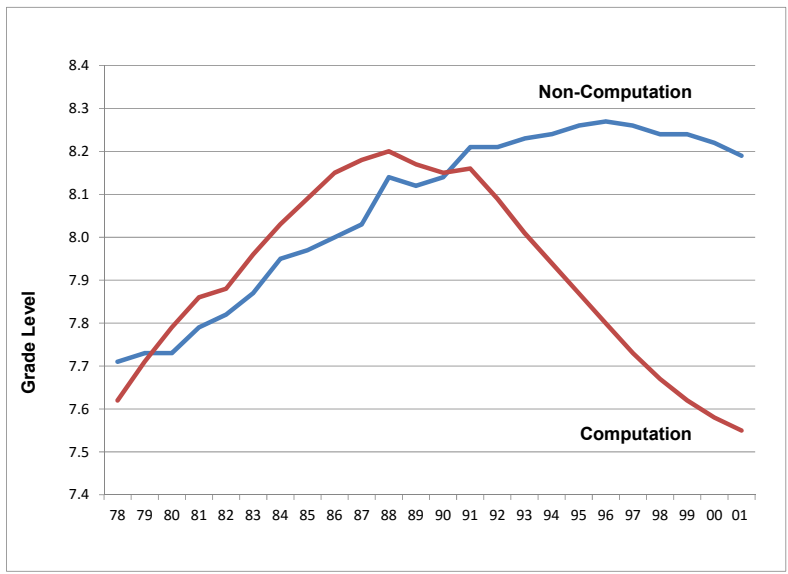

**Fig. 1 Results for 8th grade ITBS Math for Iowa, 1978-2001**

To acquire additional data on student computation skills, for this paper the authors mailed a request to state education departments in all 50 states for nationally normed math data for any years since 1990 that included a subtest in computation. Though responses are generally not required by law to information requests from non-state residents, responses were received from 29 states. Computation data were obtained from two states: Virginia and North Dakota. In all other responses, states reported that their tests were either not nationally normed or reported only a "total math" composite score. The survey results are detailed in the supplemental materials attached to the online version of this paper.

In Virginia, K-12 districts purchase math texts from a list of options provided by the state board of education. State selection is based on alignment with state K-12 math standards initially adopted in 1995 (Virginia Dept. of Education 2012). From 1998 to 2002, all Virginia local districts were required to administer *the Stanford Achievement Test, Ninth Edition* (Stanford 9) in grades 4, 6, and 9. Nearly all of Virginia's 80,000 9th graders were tested each year. Similar to the ITBS, the Stanford 9 is one of several commercially available nationally-normed tests. Norms for the Stanford 9, determined in 1995, provided a fixed standard for tests and subtests that reports results as percentiles (Virginia Dept. of Education 2003).

In "total mathematics," average Virginia scores were steady and above the national average 50th percentile (Table 2, Figure 2), but "total mathematics" scores were a composite of two subtests. For the "full" version of the Stanford 9 given during this period, as the state test report noted:

> *Two mathematics subtests are administered -- Mathematics: Problem Solving, which focuses on reasoning skills, and Mathematics: Procedures, which measures the student's facility with computation* (Virginia Dept. of Education 2003).

Calculator use was not allowed on the Mathematics: Procedures subtest (Brooks 2003).

In math reasoning from 1998 to 2002, Virginia 9th graders scored above the national average and were higher every year, while in computation, scores were below the national average and went lower every year.

**Table 2.  Stanford 9 percentile ranks for Virginia, grade 9[1]**

|                          | 1998 | 1999 | 2000 | 2001 | 2002 |
|--------------------------|------|------|------|------|------|
| Total math               | 54   | 55   | 55   | 55   | 55   |
| Problem solving (Reasoning) | 58 | 61   | 63   | 64   | 65   |
| Procedures (Computation) | 46   | 44   | 42   | 41   | 39   |

[1]Virginia Dept. of Education (2003).

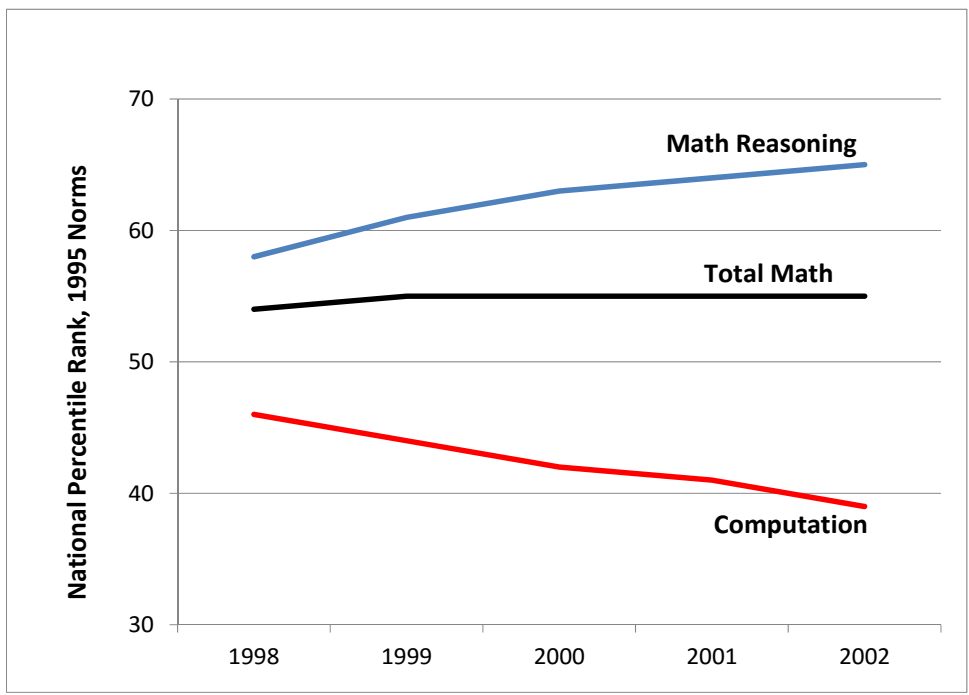

**Fig. 2** Stanford math percentile ranks for Virginia, grade 9 (Virginia Dept. of Education, 2003)

For North Dakota, test results are discussed in detail in the supplemental materials.  In brief, between 1990 and 1997 (earlier than the data for Virginia), on a different nationally normed test with fixed 1988 norms, national percentile rankings in computation were higher than in Virginia but the trend was similar:  Scores in "concepts and application" rose but in "computation" declined (North Dakota Dept. of Education, 2013).

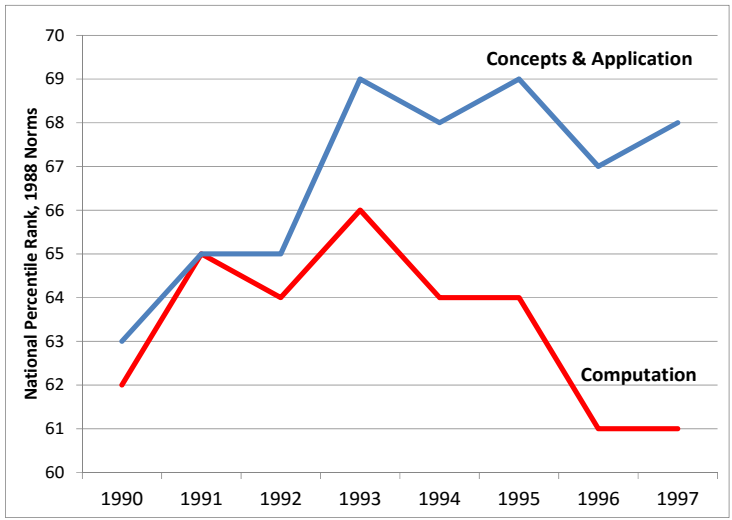

**Fig. 3  Percentiles for 8th grade math in the North Dakota CTBS/4**

## NAEP Results

The NAEP Long-Term Trend test (NAEP LTT) is the second of two US NAEP math tests administered to a national sample of students.  In contrast to the "main NAEP," the LTT has generally been scheduled every four years rather than every two, and the LTT does not permit calculator use.  The structure of the LTT was altered in 2003 but did not change between 1971 and 1999 (Beaton et al. 2011).  In 2004, Loveless and Coughlan reported that on LTT questions about fractions, scores for 17 year olds declined from a national average of 76% correct in 1990 to 56% correct in 1999 (Figure 4).

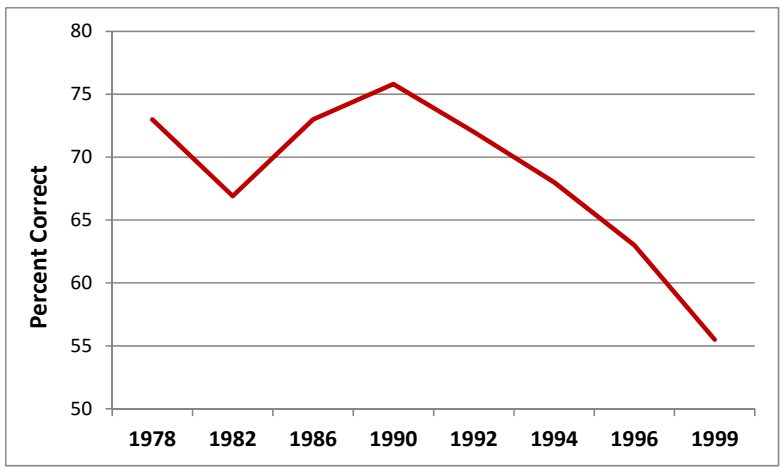

**Fig. 4  Item analysis for fractions in NAEP LTT, 1978-1999**

Fractions are encountered frequently in science calculations.  An example typical of a first-year chemistry text calculation is shown in Figure 5.

$$M = \frac{dRT}{P}$$

$$= \frac{(1.72 \text{ g/L})(0.0821 \text{ L·atm/mol·K})(298 \text{ K})}{\left(\frac{725}{760}\right) \text{atm}}$$

$$= 44.1 \text{ g/mol}$$

**Fig. 5  A Worked Example with A Complex Fraction**

A calculator can help with the numbers, but in the authors' experience, if students have not mastered fraction fundamentals, they are mystified by the unit cancellation that is essential for calculations in the sciences.

Since 1990, as noted earlier, scores on the "main NAEP" have steadily increased. However, on the age 17 NAEP LTT, scores since 1990 have seen minimal change (Beaton et al. 2011), and on the LTT in fractional computation between 1990 and 1999, scores declined substantially. What might explain the differences in the two NAEP assessments? Loveless and Coughlan note that since 1990, the content of the "main NAEP" has changed over time to reflect "the recommendations of groups advocating curriculum reform." Those changes allowed the use of calculators on parts of the main NAEP and deemphasized "shopkeeper arithmetic." In contrast, from 1971 to 1999 the NAEP LTT did not allow calculator use and retained the same structure (Loveless 2002). As one outcome, the main NAEP showed gains between 1990 and 1999 that were 10 times larger than gains on the LTT during the same period (Loveless and Coughlan 2004).

## Data 2003 To 2010

All of the data from state testing cited so far has stopped before 2003. In 2002, the federal "No Child Left Behind Act" (NCLB) supplied economic incentives for states to test each year in grades 3-8 based on state K-12 standards. By 2003, every state was conducting the tests called for by NCLB (Klein et al. 2005). Tests versus explicit standards are termed "criterion-referenced tests." From 2003 to 2013, state results typically were reported as shown in Table 3:

**Table 3:  Pennsylvania state test results for all students in the 11[th] grade**

| | Participation rate | MATH:  Percentage of students in each performance level | | | |
|---|---|---|---|---|---|
| | | Below basic | Basic | Proficient | Advanced |
| 2011-12 | 94% | 24% | 28% | 43% | 5% |
| 2010-11 | 95% | 21% | 24% | 52% | 3% |
| 2009-10 | 95% | 20% | 29% | 46% | 5% |
| 2008-09 | 97% | 27% | 28% | 40% | 5% |

[2]Pennsylvania Dept. of Education (2013). *Pennsylvania State Report Cards, 2011-12, 2009-10.*

In testing from 2003 to 2010, nearly every state had unique standards, tests, and grading scales. With nearly 50 different measures, direct comparisons could not be made among states or to national averages and no comparison can be made to historical performance.

NCLB did not require an end to norm-referenced tests, but in the responses to our survey, only a few states reported administering a nationally-normed test since 2002, and in all cases, these "limited" test versions reported "total math" but not computation subtest scores. Between 2003 and 2010, some states reported scores in "numeracy skills" on their criterion-referenced tests, but because each state test was unique, there was no standard that could be applied for comparisons to a fixed standard, other states, or national averages.

Despite the fact that spending on state testing programs was estimated in 2012 to be $1.7 billion per year (Chingos 2012), we were able to find no state that between 2003 and 2010 measured computation skills versus a viable standard.

To summarize the 1990-2010 testing results: Between 1990 and 2002, in all of the responding states that measured computation skills in any grades 8-12, test scores declined. Scores in "reasoning" and "concepts," in all cases they were reported, ranged from not correlated to negatively correlated with scores in computation. The available data indicate that scores in "total math" masked declines in computation, but "total math" scores were often the only measure reported by states. Between 2002 and 2010, no state reported test scores in a way that allowed a direct comparison to an accepted standard, other states, or national averages.

## Factors Affecting Achievement

What factors may have contributed to the decline in computation skills observed in the test score data above prior to 2003? One change cited by observers has been the increased use of calculators. The first handheld calculator was marketed in 1970, and basic calculators became widely available by 1976 for under $10 (Rapoport 1976).

Estimates of K-12 calculator use by students vary according to whether use was optional or required and which subjects and grade levels were surveyed. A 1986 NAEP survey found that 26 percent of US high school students had calculators available in math classes. By 1996 nearly 60% of 8th graders reported using calculators in math class on a daily basis (Waits and Demana 2000).

A second factor often cited as affecting computation skills was the K-12 math standards adopted in nearly every state between 1990 and 2002 (Loveless 2002, Klein 2002, Schoenfeld 2004). The National Council of Teachers of Mathematics (NCTM) is an organization composed primarily of university school of education faculty and K-12 curriculum specialists and instructors. In a 1980 position paper, the NCTM called for increased K-12 use of calculators. A more detailed 1989 position paper became known as the "NCTM Standards." Following the 1989 "Charlottesville Education Summit," between 1990 and 1997, over 45 states adopted "standards for K-12 education" (Raimi and Braden 1998), and Loveless reported in 2003: "More than forty states model their state math standards on those of the NCTM."

The NCTM standards called for the use of calculators "at all grade levels" (NCTM 1989). This recommendation has had lasting impact. In 2005, in over 30 states, math standards directed that students be taught to use calculators to solve arithmetic problems in 3$^{rd}$ grade or earlier (Klein et al. 2005). As one example, in a 2006 "Q and A" for teachers on state standards, the New Jersey Department of Education advised that:

> *"Calculators can and should be used at all grade levels to enhance student understanding of mathematical concepts. The majority of questions on New Jersey's new third- and fourth-grade assessments in mathematics will assume student access to at least a four-function calculator. Students … should be prepared to use calculators by regularly using those calculators in their instructional programs."* (New Jersey Dept. of Education 2006)

In grade 5-8 math, the NCTM standards advocated "increased attention" to "reasoning inductively and deductively" and for "creating algorithms and procedures" by students. Recommended for "decreased attention" were "memorizing rules and algorithms," "finding exact forms of answers," "manipulating symbols," "relying on outside authority (teacher or answer key)," "rote practice," and "paper and pencil fraction computation" (NCTM 1989). As most states adopted NCTM-type standards, the resulting debate on instruction became known as the "math wars" (Schoenfeld 2004, Klein 2002).

Between 2003 and 2010, findings of recent cognitive research on working and long-term memory entered the debate on math instruction. Those studies have summarized in previous publications (Hartman and Nelson 2015). Below we will review a portion of those findings that impact teaching and learning of the math pre-requisite for physics and chemistry.

## Cognitive Science

Cognitive science is the study of how the brain works and in particular how it learns. Since 1990, with the advent of new technologies to study the brain, scientific understanding of cognition has advanced rapidly (Lyon 1998).

Cognitive science divides problems into two types. "Ill-structured problems" have answers to which experts may disagree, as occurs political science and economics, while "well-structured" problems are rule-based, can be solved with structured procedures, and have precise answers (Simon 1973). Nearly all of the examples and problems in math and "physical science for science major" textbooks are well-structured.

Since 2010, nearly all cognitive scientists are in agreement on major components of a model for how non-experts (including undergraduates) should prepare to solve well-structured problems although debate continues over how students should be taught to approach ill-structured problems (Tobias and Duffy 2009, Klahr 2012). For the remainder of this paper, our discussion will be limited to the model for well-structured problem solving which applies to students between about age 12 and graduate study.

### Language Acquisition and Memory

The cognitive science model for reasoning is based on the interaction between a long-term memory (LTM) where elements of knowledge are organized and a working memory (WM) where elements are processed (Willingham 2009).

Evolution has given human children a powerful drive to learn speech. Until puberty, by unconscious processes, a child adds to their LTM words (small groups of sound elements) that they hear repeatedly.  If a word is heard with substantial frequency in different contexts, the child will add the word and the rules for its use to their spoken vocabulary (Pinker 1999, 2007, Nagy et al.1985).  By age 12, during speech, the brain applies to tens thousands of memorized words thousands of grammatical rules which children and most adults have little to no ability to state, and it does so fluently:  Quickly, effortlessly, and nearly always correctly by the rules for their dialect.  Fluency in correctly applying rules that an individual cannot explain is termed "implicit" (as opposed to explicit) understanding (Pinker 1999).

**Elements of Knowledge**

Unlike the ability to learn a language, humans did not evolve to instinctively learn reading, writing, and solving scientific calculations, but these and other "secondary" skills can be learned with focused attention, effortful practice, and instruction (Geary 2002, Sweller 2008).  Procedures (sequenced steps for processing) and facts are stored as small elements of knowledge in LTM.  As elements are linked to associated elements by their characteristics and rules for use, "conceptual frameworks" (each termed a schema, plural schemata) are constructed in LTM that convey understanding.  Human LTM has enormous capacity to hold facts, procedures, associations and rules for use, but only limited amounts of information can be stored and linked within LTM each day (Biemiller 2001; Sweller 2009).

The brain thinks, plans, and solves problems in working memory (WM).  WM can accept input both from the environment via the senses and from LTM (Clark et al. 2012).   Humans solve problems by a process similar to how we construct speech:  By moving into WM and then processing small elements of knowledge.  Elements in WM serve as cues that promote retrieval from LTM into WM of associated elements that in similar contexts have previously been used in problem solving (Willingham 2009b; Anderson and Lebiere 1998; Anderson et al. 2000).

**Limits in Working Memory**

WM has an essentially unlimited ability to hold and process elements "memorized to automaticity" so that in response to cues they can quickly and accurately be recalled from LTM (Ericsson and Kintsch 1995).  However, WM can retain a "non-automated" element only 30 seconds or less unless it is rehearsed (Peterson and Peterson 1959), and at any given point during processing can hold only about 3-5 elements that are "novel," meaning they cannot be automatically recalled from LTM (Cowan 2001, 2010, Clark et al. 2012).  Two ways to state the implications of WM characteristics are:

- *Working memory (where you think) can recall unlimited well-memorized information but can hold only a few small elements that are not-well-memorized,* and
- *Your ability to solve problems depends nearly entirely on how much knowledge you have "memorized to automaticity" in LTM.*

**Overcoming WM Limits**

During problem solving, if the limits in novel WM are reached, the result is a sense of confusion and a likely inability to solve (Clark et al. 2012).  Three ways around the novel WM constraints are chunking, algorithms, and automaticity.  All require thorough memorization.

"Chunks" are elements that have been memorized as a group.  For example, for most Americans, the 9 letters YTQRMZAXH will likely be more difficult to remember than the 15 letters of NBCABCCBSCNNFOX (Willingham 2006).

Algorithms are a sequence of problem solving steps that limiting how many non-memorized elements must be held in WM at any point during processing.  Memorized algorithms combine the goal and steps into one recallable chunk and minimize the impact of the "30 second or less" limit for holding novel elements during steps of problem solving (Willingham 2004, 2006, 2008).

## Automaticity

What cognitive scientists call the "central" strategy to improve student problem solving is "memorization to automaticity."   In the 2008 *Report of the National Mathematics Advisory Panel* (NMAP), cognitive experts Geary, Reyna, Siegler, Embretson, and Boykin write,

> "[T]here are several ways to improve the functional capacity of working memory. The most central of these is the achievement of automaticity, that is, the fast, implicit, and automatic retrieval of a fact or a procedure from long-term memory (Geary et al. 2008).

Most instructors will likely agree that a student with a poor math background will tend to have difficulty in quantitative science courses.  What may be surprising is *how well* students must memorize fundamentals to be adequately prepared for scientific problem solving.  The NMAP authors add,

> "[During calculations,] to obtain the maximal benefits of automaticity in support of complex problem solving, arithmetic facts and fundamental algorithms should be thoroughly mastered, and indeed, over-learned, rather than merely learned to a moderate degree of proficiency (Geary et al. 2008).

"Over-learned" means that facts and fundamental algorithms are memorized until they can be recalled fast, perfectly, and repeatedly.  The rules for mathematics also apply when solving well-structured problems in the physical sciences (Anderson et al. 2000).

For topics in which an individual is not an expert, adding new knowledge to LTM requires slow physiological changes as the brain is in a sense "re-wired." (Clark 2006, Willingham 2004, 2006). Assimilation is promoted by "self-testing" (such as "clicker questions" and flashcard drill) and "spaced overlearning" (recall mastery achieved repeatedly over several days) (Brown 2014, Willingham 2002, 2004, 2008).

To recall elements at appropriate times, practice is necessary that applies new elements in different contexts.  Novel information processed in WM also enters LTM where it will "stick" eventually if it is processed frequently -- or quickly if it "fits" in an existing schema.  If space is available in novel WM during processing, an element and its context elements are moved together into LTM, and their linkage strengthens conceptual frameworks.  Brown et al. (2014) document the value of *interleaved practice, reflection, elaboration*, and *faded guidance* (scaffolding) to develop an intuitive sense based on context of which facts and algorithms should be recalled to solve a problem.  When a range of fundamental elements pertaining to a topic can be recalled automatically and at varied appropriate times, the result is termed fluency.  Once information is well memorized and well organized in LTM, the ability of WM to recall and apply that knowledge fluently often lasts for decades.

 "The wisdom of using minimally guided techniques to reinforce or practice previously learned material" is noted by Clark et al. (2012).  As active learning tags new knowledge with visual, auditory, and semantic associations, linkages in LTM both generalize and differentiate meaning (Willingham 2003, 2008). Instructor-guided inquiry can also motivate students to persevere in the retrieval practice necessary to achieve automaticity (Abraham 2005; Duckworth et al. 2007).

**Is Learning to Calculate Necessary?**

In an age when computers are automating the solving of many problems, must students put in the time to automate the steps to solve those problems without technology? Cognitive experts respond that, even in the age of computers, "automaticity in support of complex problem solving" (see NMAP above) is crucial for students because "complex problems have simple problems embedded in them" (Willingham 2009b, Clark 2010). If students are presented with a complex stepwise calculation that computers are not programmed to solve and steps to solve components have not previously been automated, slots are unlikely to be available in novel WM for the management of multiple problem-solving steps (Anderson et al. 2000).

As one example, if as part of a calculation "8 times 7" cannot be recalled, the calculator 56 must be stored in WM so that it can be transferred to where the calculation is being written. On a problem of any complexity, that storage may bump out of the 3-5 novel WM slots an element that is needed to solve the problem. An answer from a calculator takes up limited working memory space; an answer recalled from long term memory does not.

If arithmetic and algebraic fundamentals are automated, when examples are based on simple ratios or equations, room is available in novel WM for the context that builds conceptual understanding, and problem solving builds an intuitive, fluent understanding of when to apply facts and procedures (Willingham 2006).

Conversely, if a student lacks "mental math" automaticity, conceptual explanations based on proportional reasoning or "simple whole-number-mole ratios" will likely not be simple. If a student must slowly reason their way through steps of algebra that could be performed quickly if automated, the "30 seconds or less" limit on holding the goal, steps, and data elements of the problem in WM ticks away.

In recent years, the internet has facilitated the finding of facts and procedures, but new information occupies the limited space in novel WM that is needed to process the unique elements of a problem. Unless new information is moved into LTM by repeated practice at recall, during future problem solving that new information will again need to be sought, and when found, it will again restrict cognitive processing (Willingham 2004, 2006).

**The Limits of Reasoning**

Does improved reasoning skill, by itself, help students solve problems? The Virginia data (Figure 2) say no, and cognitive studies explain why. Research has verified that reasoning and "critical thinking" strategies are either very general (and usually well understood by most individuals) or domain specific (Sweller et al. 2010). Domain strategies can be taught relatively quickly, while automation and building frameworks in a domain are the slow and therefore rate-determining steps in learning (Willingham 2007). Reasoning strategies are important, but not sufficient. For novice learners, having to reason, test, and hold the steps of a novel procedure will quickly overwhelm the limits of novel WM.

The Virginia test results are consistent with the cognitive finding that, except for a small number of data elements for a problem, students can only reason with what they reliably remember, they remember only facts and procedures they have practiced recalling and applying (Willingham 2003, 2007). Virginia's standards, adopted in 1995, directed teachers to prepare students to use calculators after 3rd grade (Virginia Dept. of Education 1995). When practice in mentally recalling and applying computation facts is de-emphasized, science predicts computation abilities will decline, consistent with the observed Virginia results.

This is not meant to imply that math reasoning is not important. However, to solve scientific calculations, computation fluency is also essential. Because math reasoning and computation skills are not necessarily correlated (Figure 2), to measure a student's readiness to solve calculations in the sciences, fluency in computation must be evaluated separately from reasoning.

## Standards and Achievement

Since 1995, preparation for tests on state standards has become a focus of K-12 instruction. Between 1990 and 2002, over 40 states adopted NCTM-type math standards, and publishers increasingly marketed NCTM-aligned texts (Loveless 2003). Those standards directed teachers to de-emphasize recalling math facts from memory, "memorizing rules and algorithms," "finding exact forms of answers," "manipulating symbols," "rote practice," and "paper and pencil fraction computation," and most states retained NCTM-type standards until 2010 (Klein et al. 2005, Carmichael et al. 2010). Recent cognitive research suggests that students taught under such standards would tend to have difficulty solving calculations, but do data support this prediction?

In the data available between 1990 and 2002, computation scores declined. Between 2003 and 2010, little to no measurements of computation skills was reported from tests on state standards, but other measures are available. Geary et al. (2008) note that between 1991 and 2001, studies of US college students found that many could not recall basic addition, subtraction, and multiplication facts. Larger data samples for the nation and some states were gathered in 2011 and 2012 by the TIMSS and PIAAC testing.

### TIMSS Results

The *Trends in International Mathematics and Science Study* (TIMSS) is an international test administered in 1995, 1999, 2003, 2007, and 2011 in 38 to 57 "educational systems" including the US. Most results were reported by nation, but scores were also reported for some US states, Canadian provinces, and large Asian cities. Scores are "re-centered" to 500 in each year of administration, with 600 or 400 representing one standard deviation from the mean. TIMSS is given in 4th and 8th grade. Between 1995 and 2011, in "overall 8th grade mathematics," the US average score ranged between 500 and 509 (Provasnik et al. 2012, National Center for Education Statistics 2015).

The "overall" 2011 TIMSS score is a composite of four subtests: Number (30%), algebra (30%), geometry (20%), and data/chance (20%). In 2011, scores on each subtest were reported. On the 8th grade TIMSS, the "number" content domain is defined as "understandings and skills related to whole numbers, fractions and decimals, integers, ratio, proportion, and percent." Calculator use is optional and is decided by each nation, but problems are designed to be done without a calculator (Mullis et al. 2014). Scores in the number domain are the least likely to be influenced in 8th grade by the different sequences in which systems cover topics. In "number," the US 2011 average was 514 (Table 4).

US states may volunteer to have their TIMSS scores reported.  In 2003, data were reported for one state (Indiana), in 2007 for two states (Minnesota and Massachusetts), and in 2011 for nine states (Table 4).

Between 1997 and 2003, four US states (Massachusetts, Minnesota, Indiana, and California) shifted from NCTM-type standards toward an emphasis on arithmetic automaticity, and in 2011 each of those states voluntarily participated in TIMSS reporting.  Data for those states provide some evidence of the impact of state standards on the math automaticity needed to solve scientific calculations.

In 2000, Massachusetts shifted standards "toward more emphasis on computational facility" (Schmid 2000, Riordan and Noyce 2001).  For the 2011 TIMSS, in overall 8th grade math, Massachusetts ranked highest among the 9 TIMSS-reporting US states, scoring 53 points above the US average, and scored just below Japan (Table 4).  On the "number" subtest, the Massachusetts score of 567 was 10 points higher than Japan's.

Minnesota in 2003 changed to a policy of "no calculators on state elementary math tests" (National Council on Teacher Quality 2009).  On the 2011 8th grade TIMSS, Minnesota's ranked 2nd among reporting US states and on the "number" subtest scored one point below Japan.

**Table 4: TIMSS National and State Data, Partial Listing**

| TIMSS 2011 – 8th Grade Math | | |
|---|---|---|
| Education System | Overall Mathematics Score | Number Subtest |
| South Korea | 613 | 618 |
| Singapore | 611 | 611 |
| Japan | 570 | 557 |
| *Massachusetts US* | 561 | 567 |
| *Minnesota US* | 545 | 556 |
| Russian Federation | 539 | 534 |
| *North Carolina US* | 537 | 547 |
| Quebec | 532 | 543 |
| *Indiana US* | 522 | 528 |
| *Connecticut US* | 518 | 527 |
| *Colorado US* | 518 | 521 |
| Israel | 516 | 518 |
| Finland | 514 | 527 |
| *Florida US* | 513 | 517 |
| United States | 509 | 514 |
| England | 507 | 512 |
| Italy | 498 | 496 |
| *California US* | 493 | 492 |
| Kazakhstan | 487 | 479 |
| Sweden | 484 | 504 |
| *Alabama US* | 466 | 463 |

Indiana revised standards in 2000 to emphasize mental calculations and standard algorithms (Klein et al. 2005). On the 2011 TIMSS, Indiana ranked 4th among the 9 reporting states in "overall mathematics" and "number" and 14 points above the US average in both overall math and the number subtest.

TIMSS state data show a wide variation even for states in the same region: On the number subtest, North Carolina averaged 547 and Alabama 463. Between 1998 to 2003, North Carolina standards emphasized arithmetic fluency, though standards after 2003 adopted less specific guidance on fluency and calculator use (Raimi and Braden 1998, Klein et al. 2005).

California revised standards in 1997 to emphasize fluency in arithmetic. Of the four states that by 2003 had adjusted standards to support automaticity in "mental math," only California scored below the national average on the 2011 8th grade TIMSS. One factor may be that most K-12 districts in California operated under significantly reduced budgets in the years following the 2000 "dot-com" and 2007 mortgage crises (California Dept. of Education 2013). California may be a reminder that funding for new curricula, professional development, class size, and instructor retention also impacts K-12 student performance. Details on the California data are discussed in the Supplemental Materials.

We could find no data for states or the nation that specifically measure algebraic fluency at or near the end of high school. Overall for the US, the 2011 TIMSS scores of 509 in "overall mathematics" and 514 in "number" are close to average of 500 for over 40 nations tested, nearly all of which are far more limited in economic resources than the US.

**PIAAC Data**

In February 2015, an additional measure of US student preparation for physical science and engineering courses was reported by the US-based Educational Testing Service (ETS). The ETS analyzed data from the 2012 Programme for the International Assessment of Adult Competencies (PIAAC), a survey by the Organization for Economic Co-operation and Development (OECD) of 22 OECD member-nations (all highly developed economies). PIAAC reports results in three categories: Literacy, numeracy, and "problem solving in technology-rich environments" (PS-TRE), which assesses how well participants "interact effectively with digital technology" during problem solving. Data are reported by age groups between 16 and 65. Calculator use was permitted on the numeracy subtest (Gal et al. 2009). The numeracy tested was defined as "solving a problem in a real context by responding to mathematical content/information/ideas represented in multiple ways."

ETS noted that "researchers have found that numeracy is a robust predictor of labor market success." Among participating nations, the youngest US cohort (ages 16 to 24) was found to be "ranked dead last in numeracy and among the bottom in PS-TRE" (Goodman et al. 2015).

# Common Core Standards

During 2010-2011, 44 states the adopted new K-12 "Common Core" standards (Common Core State Standards Initiative 2010, Carmichael et al. 2010). Some states later reversed adoption, but in December 2014 the Common Core remained in place in 39 states. To evaluate student performance versus the standards, two separate consortia are developing tests that began in 2015 (Gewertz 2015).

The Common Core supports fluency in some arithmetic fundamentals, a change from standards in most states between 1995 and 2010. On the tests from both of the consortia, students are not allowed to use calculators until grade 6 (Robelen 2013). However, the standards call for fluency in only a limited number of topics in algebra.

In addition, the new standards call for explicit understanding of mathematical rules (being able to explain why). Learning explicit rules is a time-consuming process, and cognitive science has found that for non-expert learners, reasoning from explicit rules without fluent (implicit) recall of algorithms is not likely to lead to problem solving success. Fluent recall of facts and algorithms does result in high success rates, but extensive practice is required to develop this implicit understanding (Clark et al. 2012, Geary et al. 2008). Whether students will be able to achieve both implicit recall and explicit understanding in the K-12 time allotted for math is as yet unknown.

For states using a consortium test, comparisons to national averages will be possible for subtests reported, but more than half of US students were scheduled in 2015 to take a test unique to their state (Gewertz 2015). When a test is unique, comparing student achievement to any viable standard is difficult at best.

**Will Students Be Prepared?**

At this point, it is not possible to predict the Common Core impact, but in preparing students for scientific calculations, even ideal standards would only slowly repair deficits. After purchase, K-12 text series are typically loaned to students each year for the next six years, so that editions purchased in 2011, before they could have been Common Core aligned, would be the basis for K-12 curriculum until 2017. Students from 2017 first grade classes will arrive in college in 2029, and without special attention, those who arrive through 2029 may be disadvantaged by texts that during early schooling discouraged arithmetic automaticity.

# Conclusion

For the past two decades, a dominant narrative in K-12 public education has been that students are not learning as much as they should and the reason has been poor teaching. To force changes, remedies implemented have included vouchers, charter schools, reconstitution (replacing faculty at schools), and state tests used to determine teacher evaluation and pay. As the Virginia, Iowa, and LTT data graphically illustrate, there is evidence that student computation skills needed for scientific calculations did indeed fall precipitously, but they did so only after textbooks and/or state standards required teachers to de-emphasize math automaticity. TIMSS state data suggest that, in the absence of sharp reductions in financial support for education, a reform that has measurably and significantly improved student computation achievement has been to supply teachers with math standards that take into account how the brain works, plus training in how to implement them (Schmid 2000, National Council on Teacher Quality 2009).

This is not to say that the adoption of standards that discouraged automaticity was the "fault" of any individuals or groups. In the 1990's, as K-12 standards were initially being adopted, it was widely assumed that the brain could hold and process newly acquired information from calculators, computers, or recent reasoning as easily as it processed thoroughly memorized information. Only gradually between 2000 and 2010 did research verify that for well-structured problem solving, that assumption was incorrect.

For this study, our first question was, "In physical science courses, is it important that students be able to solve fundamental mathematical calculations without reliance on computers or calculators?" We found that science says yes. "Arithmetic facts and fundamental algorithms" must be "thoroughly mastered, and indeed, overlearned" to avoid the bottleneck in novel working memory.

Question 2 asked: "In the US, how much has student math background been a factor in low success rates in introductory physical science courses and the decline in degrees in the physical sciences and engineering?" We cannot measure precisely the extent to which low STEM success has been due to computation deficits, calculator use, K-12 standards, school budgets, changing teacher or student characteristics, or other factors. However, test data generally indicate, as cognitive science predicts, that states discouraging automaticity saw computation skills decline, but most states adopting standards favoring arithmetic automaticity saw calculation skills rise to high levels compared to other states and most nations. These data also indicate that in some cases, standards alone did not predict math outcomes.

Question 3 asked: "Can we identify actions that may be the most promising in improving success in quantitative STEM courses?" Without experiments, we do not know how much improved math automaticity might improve STEM success, but we do know that in 2015, students entering college from most states are likely to have spent a number of K-12 years with standards and textbooks that did not encourage arithmetic and algebraic fluency.

Cognitive science suggests that if pre-requisite math has not previously been automated, a student will find calculations in the physical sciences to be quite difficult, but that if automaticity deficits are addressed before a math topic is needed for a topic, success rates should increase. To determine whether fluency deficits may be a significant restraint for a particular student population, a brief quiz could be given to a sample of students. An editable 15-minute quiz on "math automaticity" designed by the authors for small or recitation sections of introductory physics or chemistry is posted at  http://bit.ly/1HyamPc . For larger sections, a editable multiple choice quiz in "calculator-free math fluency" is available from Leopold and Edgar (2008).

If a brief quiz shows that arithmetic or algebraic deficits are significant, a placement test that measures math automaticity could be considered. If a "computation-specific" test is not practical, a "calculator-free math fluency" section might be included in a broader math placement test, with separate reporting of those scores. Testing each year would track the gradual impact of the new K-12 math standards being implemented in most states.

For first-year college undergraduates found to have computation deficits, a 1-3 credit course could be offered in "math for the physical sciences" prior to, or concurrent with, introductory physical science courses. Whether a one-semester course would help to a significant number of students will vary depending on student backgrounds, but given the importance to our society of success in STEM study, a variety of experiments to improve math automaticity may be worthwhile to explore. Such a course should not be considered "remedial" for students from states where, between 1995 and 2010, standards de-emphasized automaticity in computation skills. A topic is not remedial if it has not previously been a goal of instruction.

During high school, if an initial "mental math" quiz reveals deficits, review and practice could be provided prior to the point a topic is needed in introductory physics or chemistry courses. For students already "in the K-12 pipeline," instructional materials specially designed to build fluency could be developed for teacher use. Instructional technology may assist with materials that add interest to the practice required to achieve automaticity. Thanks to scientific progress in our understanding of cognition, in 2015 we have better understanding of specific instructional practices that can help make study "stick."

If a portion of state math tests through the end of high school evaluate "mental math" skills, it would encourage STEM readiness. This would not require more K-12 testing, but simply that the components of math important in scientific calculations be included on tests and reported in scores. What gets reported gets taught.

In large measure, quality of life in a modern society is determined by the ability of its workforce to apply mathematics to solve scientific and technical problems. Historically, a central purpose of K-12 instruction was to develop student fluency in arithmetic and algebra needed to solve technical calculations, but as TIMSS and PIAAC data indicate, numeracy skills for US students have fallen far behind competing nations. Cognitive studies indicate that a significant factor in this decline has been K-12 math standards (decided above the instructor, school, and district level) that later proved to be misaligned with how the brain solves problems.

Recent research has increased scientific understanding of what the human brain can and cannot do. That research offers guidance in improved strategies to nurture student achievement in science, mathematics, and technology. If instructional standards and policies take this scientific progress into account, our students and our society will benefit.

### Supporting Information:

The file of supporting information below the references contains:

1. Discussion of the nationally-normed test score data for North Dakota, 1990-2000.

2. Discussion of the math standards and test results in California.

3. Survey of State Departments of Education and Response Summary.

**Conflict of Interest.** The authors declare the following competing financial interests: Eric Nelson is a co-author of textbooks in chemistry.

*Title*:

# Automaticity in Computation and Student Success in Introductory Physical Science Courses

*Authors*:

JudithAnn R. Hartman,* Eric A. Nelson[†]

*Author Affiliations:*

*Department of Chemistry, United States Naval Academy, Annapolis, MD 21042 USA

[†]Fairfax County (VA, USA) Public Schools (retired)

*Corresponding Author:*

 *Hartman@usna.edu

*Contents:*

This supplement addresses these topics:

1. Nationally-normed test score data for North Dakota, 1990-2000.
2. Detail on the math standards and test results in California.
3. Survey of State Departments of Education and Response Summary

**1. North Dakota: 1990-2000**

In response to our survey, the North Dakota provided data on nationally normed testing that included a subtest in math computation.

From 1990 to 1997, North Dakota administered to nearly all state 8[th] graders, approximately 8,000 students each year, the Comprehensive Test of Basic Skills, Fourth Edition (CTBS/4) based in 1988 norms. From 1998 to 2000, North Dakota administered the CTBS/5, also known as the Terra Nova, based on 1996 norms. For both versions of the CTBS, an overall score was reported in "total mathematics," and subtest scores were reported in "computation" and "concepts and application."

Data from the CTBS/4 is included in the main section of this report. Because the CTBS/5 is a different test based on a different norming sample; results from the two assessments are not comparable. For the CTBS/5, data for the three available years 1998-2000 showed no trend (Table 3).

|  | 1998 | 1999 | 2000 |
|---|---|---|---|
| Computation | 67 | 67 | 67 |
| Concepts & Application | 65 | 67 | 65 |
| Mathematics Total | 67 | 68 | 67 |

**Table 3.  North Dakota-Grade 8 Math, CTBS/5, National Percentile Ranks**

North Dakota also administered the CTBS/4 in 11[th] grade and CTBS/5 in 10th grade during the 1990-2000 period. Computation shows a small rise and then fall from 1990 to 1997; concepts show a small but reasonably steady increase during this period (Table 4). Trends are similar to data for Iowa, Virginia, and 8[th] grade North Dakota, but are small in magnitude.
The change in the test between 1997 and 1998 renders the longitudinal study for North Dakota more difficult during the period when the changes noted in Iowa and Virginia were most apparent.

|  | 1990 | 1991 | 1992 | 1993 | 1994 | 1995 | 1996 | 1997 | 1998 | 1999 | 2000 |
|---|---|---|---|---|---|---|---|---|---|---|---|
| Grade | 11 | 11 | 11 | 11 | 11 | 11 | 11 | 11 | 10 | 10 | 10 |
|  | CTBS/4 | | | | | | | | CTBS/5 | | |
| Computation | 63 | 64 | 66 | 67 | 67 | 65 | 65 | 64 | 66 | 68 | 67 |
| Concepts & Application | 64 | 65 | 66 | 67 | 68 | 67 | 68 | 69 | 72 | 74 | 73 |
| Mathematics Total | 64 | 66 | 67 | 68 | 68 | 67 | 68 | 67 | 71 | 73 | 72 |

**Table 4.  North Dakota-Grade 11 Math, National Percentile Ranks**

## 2. Case Study: California

The history of the math wars in California has been detailed from multiple viewpoints [1-3]. In brief: For math textbooks in grades 1-8, California is an "adoption state." The state board of education adopts a list of approved textbooks, usually every seven years. Within a year or two of state adoption, local districts are generally required to purchase and loan to students books which the district chooses from the approved list. Compared to "non-adoption" states in which a change in the curriculum in textbooks would then be adopted gradually over 6-7 years, in California a change in curriculum direction at the state level is more quickly implemented by localities.

In California, the state board begins each adoption cycle by approving a curriculum framework and content standards. Textbook publishers are invited to prepare and submit texts that are aligned with the standards, and within three years, textbooks are chosen for the approved state list. Although the math framework covers courses through high school calculus, local districts are permitted to adopt any texts of their choosing for 9$^{th}$ grade and above. In grades K-8, local districts may also apply for a waiver to choose a non-state-adopted text [4].

A 1980 NCTM position paper, the *Agenda for Action,* advocated many of the math practices later included in the 1989 NCTM standards, including increased reliance on calculators. In 1985, California adopted a math framework which included elements of the *Agenda for Action*. In 1992, a new math framework was adopted in California that was closely aligned with the 1989 NCTM standards. [3]

In the 1996 main NAEP testing, California 4$^{th}$ grade math scores ranked 42nd out of 44 reporting states, above only Louisiana and Mississippi [5]. In the California State University (CSU) system, the percentage of first-year students needing a course in remedial math rose from 23% in 1989 to 54% in 1997. [3]

Amid growing controversy, in 1997 the state board asked a group of Stanford University math professors to draft new math standards which the board subsequently adopted. Addressing the new standards, an NCTM newsletter charged that "mathematics education in California suffered a serious blow," citing "curriculum standards that emphasize basic skills and de-emphasize creative problem solving, procedural skills, and critical thinking." [6] California's Superintendent of Schools termed the new standards "dumbed down" and "a decided shift toward less thinking and more rote memorization," noting that "we're not even going to let [students] use a calculator before the sixth grade." [7]

Supporting the 1997 standards was an open letter signed more than 100 California mathematics professors and instructors, including Jaime Escalante and the chairs of the math departments at Stanford and Cal Tech. [8] The letter's final paragraph stated, "Good mathematics standards require mastery of both basic skills and broader mathematical concepts."

Textbooks aligned with the 1997 standards were adopted by the state in January, 2001. [4] The 1997 framework remained in place with only minor changes until 2012. What was the effect of the new standards on achievement?

In 2003, California began administration of criterion-referenced tests on the California standards. Enrollment and achievement data from 2003 to 2013 are shown in the tables below. [9]

Between 2003 and 2013, overall state K-12 enrollment increased by less than 1% [10] while the number of students completing Algebra I increased by 35% and Algebra II by 80% (Table 6). However, the number of students completing Algebra II by the end of the 11[th] grade in 2013 was only 43% of the number who completed Algebra I.

| Test | 2003 | 2005 | 2007 | 2009 | 2011 | 2013 | Change in Number 2003-2013 |
|---|---|---|---|---|---|---|---|
| General Mathematics | 451,126 | 374,900 | 307,656 | 259,494 | 201,133 | 188,410 | -262,716 |
| Algebra I | 505,883 | 681,924 | 744,814 | 758,859 | 740,286 | 681,237 | 175,354 |
| Geometry | 270,560 | 333,334 | 371,118 | 399,539 | 408,354 | 405,328 | 134,768 |
| Algebra II | 162,672 | 196,079 | 231,335 | 251,251 | 277,846 | 294,094 | 131,422 |

The California Modified Assessments (CMA) end-of-course mathematics tests for students with disabilities were introduced in 2010. Approximately 36,000 students took a 2012 CMA end-of-course mathematics test and are not included in this table. [9]

**Table 6. California, Mathematics, Numbers of Students Tested** [9].

Between 2003 and 2013, overall, scores on end-of-course tests in math generally rose (Table 7). Initially, scores dropped in some math courses above grade 7, but in Algebra and above, enrollment was rising significantly during the 2003-2013 period. State textbook adoption applied only to Grades 1-7; in Algebra and above, localities were free to adopt math textbooks of their own choosing.

| Grade | 2003 | 2005 | 2007 | 2009 | 2011 | 2013 | Change in Percentage 2003–2013 |
|---|---|---|---|---|---|---|---|
| Grade 2 | 53 | 56 | 59 | 63 | 66 | 65 | 12 |
| Grade 3 | 46 | 54 | 58 | 64 | 68 | 67 | 21 |
| Grade 4 | 45 | 50 | 56 | 66 | 71 | 72 | 27 |
| Grade 5 | 35 | 44 | 49 | 57 | 63 | 65 | 30 |
| Grade 6 | 34 | 40 | 42 | 49 | 53 | 55 | 21 |
| Grade 7 | 30 | 37 | 39 | 43 | 49 | 52 | 21 |
| General Mathematics | 20 | 22 | 21 | 26 | 28 | 28 | 8 |
| Algebra I[†] | 21 | 19 | 24 | 28 | 33 | 36 | 15 |
| Geometry | 26 | 26 | 24 | 26 | 30 | 29 | 3 |
| Algebra II | 29 | 26 | 27 | 28 | 33 | 33 | 4 |

[†]Prior to 2007, Algebra I was an end-of-course test for grades 8-11 students. Beginning in 2007, students in grade seven were allowed to take the Algebra I test.

**Table 7. California, Mathematics, Percentage of Students Scoring at Proficient and Above.** [9]

In high school science, California also reports enrollment data and test results (Tables 8 and 9) in grades 9 through 11 (the omission of grade 12 may disproportionately affect the reported physics enrollments and scores).

| Test | 2003 | 2005 | 2007 | 2009 | 2011 | 2013 | Change 2003-2013 |
|---|---|---|---|---|---|---|---|
| Earth Science | 89,676 | 173,958 | 207,246 | 226,308 | 215,875 | 199,380 | 109,704 |
| Biology | 334,005 | 453,685 | 507,155 | 535,179 | 552,914 | 553,575 | 219,570 |
| Chemistry | 153,491 | 196,700 | 227,866 | 247,306 | 265,399 | 285,538 | 132,047 |
| Physics | 44,878 | 59,382 | 63,450 | 67,845 | 76,199 | 85,473 | 40,595 |
| Integrated 1 | 62,008 | 111,366 | 96,818 | 69,602 | 54,950 | 39,802 | -22,206 |
| Integrated 2 | 25,983 | 20,629 | 13,822 | 4,647 | 4,119 | 3,418 | -22,565 |

**Table 8. California, Science, Numbers of Students Tested, End-of-Course Tests, Grades Nine Through Eleven.** [9]

From 2003 to 2013, the number of students (by the end of the 11th grade) completing physics increased by 90% and chemistry increased by 86%.  Mirroring math, science gains in both enrollment and achievement became higher as students tested had spent more years under the post-1997 math framework and with post-2001 math textbooks.  Achievement in math and science courses may have been impacted as well by a shift toward phonics-based reading instruction in California during this period [11].

| Test | 2003 | 2005 | 2007 | 2009 | 2011 | 2013 | Change in Percentage 2003–2013 |
|---|---|---|---|---|---|---|---|
| Earth Science | 21 | 23 | 26 | 28 | 35 | 37 | 16 |
| Biology | 37 | 32 | 37 | 42 | 49 | 50 | 13 |
| Chemistry | 31 | 27 | 31 | 36 | 38 | 40 | 9 |
| Physics | 29 | 31 | 35 | 46 | 52 | 53 | 24 |
| Integrated 1 | 7 | 8 | 10 | 13 | 20 | 24 | 17 |
| Integrated 2 | 8 | 6 | 7 | 15 | 19 | 22 | 14 |

**Table 9.  California, Science, Percentage of Students Scoring at Proficient and Above, End-of-Course Tests, Grades Nine Through Eleven** [9]

For the rise seen in math and science test scores, especially if consequences are attached to the tests, caution is always in order.  What is the test security?  Is the test changing in difficulty over time?  Are some groups increasingly being excluded?  Did the grade level at which the courses were taught change over time?  What is being tested, and is it what is most important?

Assuming reasonable answers to those questions, California's improvements in both enrollment and achievement should mean increased student readiness for college science, but the improvements started from relatively low 2003 base.  Of concern is the low number of students reaching "proficiency" in high school math and science courses, if the proficiency definition is reasonable.

California's reporting of its 2011 TIMSS scores, testing beyond grades 3-8, and the comprehensive reporting of test results and enrollments over time, all reflect a commitment to transparency and accountability which other states might emulate.  However, the data show that, compared to 2013 algebra I enrollment, a relatively low percentage of students as of 2013 had been or were enrolled in algebra II and chemistry by the 11th grade, despite substantial increases in the number of students taking those courses compared to 2003.  In 2013, the number of students who tested in end-of-course chemistry by grade 11 is only 42% of the number who tested in end-of-course Algebra I.

**Rate of Change**

In terms of how long it can take for K-12 change to impact STEM college graduation rates, California is a cautionary tale. Until 2007, the golden state's high school math and science scores showed minimal gains. This may indicate that students who had to transition to new standards "mid-stream," on a foundation of the NCTM-type standards, had difficulty. It may also reflect the need for more extensive staff development for faculty when changes in curriculum methodologies are required, as well as financial strains on K-12 districts in California during this period.

Because the California data collection ends at the 11th grade level, analysis of data on physics, which is more likely to be taken in the 12th grade than chemistry, may omit a sizeable percentage of students. For most students, chemistry is a 10th or 11th grade course, and data through grade 11 would include most students in chemistry.

Students in 2011 in chemistry showed significant test score gains over students in 2007. One factor may be that 2011 students, if tested in chemistry at the end of the 11th grade, had started 1st grade in 2001, at about the time when math texts were adopted that were aligned with the 1997 framework. These students were not as likely to encounter gaps in preparation due to different curriculum in the old versus new standards.

The 2011 cohort with significantly improved chemistry scores would be scheduled to enroll in first-year college science in the fall of 2013 and graduate in 2017, 20 years after the adoption of California's 1997 standards. Most states adopted the common core standards in 2010, 13 years after California's adoption of standards supporting fluency. These results suggest that in other states, unless efforts are made to provide special help in computation to students who have already started K-12, STEM undergraduate graduation rates may not rise from current levels for many years after the adoption of improved standards.

**California TIMSS Results**

Of the four states that changed their math standards between 1997 and 2003 to emphasize math fluency, Massachusetts and Minnesota in 2011 did exceptionally well in TIMSS results, and Indiana scored notably above the US average in overall math and numeracy. California was the only state that did not achieve 2011 TIMSS scores above the US average.

In the 2011 TIMSS testing, California 8th graders scored 493 in "total math," compared to a US average of 509. On the numeracy subtest, the California score of 492 compared to a US average of 514 (see main article, Table 4).

Observers generally agreed that the 1997 California standards were clear, coherent, and emphasized arithmetic fluency [12]. Are there possible differences among Massachusetts, Minnesota, Indiana, and California that might explain these different outcomes?

Tax revenues in California were disproportionately reduced, compared to most states, by the 2000 "dot-com" and 2007 "mortgage" crises, and as a result, most California K-12 districts operated under significantly reduced budgets (California Dept. of Education 2013). State and local budget crises have many ramifications. Declines in funding in local districts can impact funding for items including professional development, classroom observations, and class size.

One factor affecting state TIMSS scores may be that California was an "early adopter" of NCTM-type standards in its 1985 and 1992 standards. The change toward standards favoring automaticity in 1997 therefore came after teachers and local administrators had been implementing NCTM-type standards and textbooks for a longer period than teachers in most states. The 1997 abrupt reversal from many years in curriculum direction to another, during a period when limited funds were available for staff development, may have meant that the standards changed more than actual instructional practices in the classroom. School districts may have purchased new texts that aligned with the 1997 standards, but teachers may have retained for classroom use texts aligned with the older, more familiar standards. As a result, the 1997 California math standards favoring arithmetic automaticity may have been implemented in some classrooms at a very slow rate.

Another explanation may be that the California 2011 TIMSS scores represented substantial improvement over where the state had been 10 years earlier. This would be consistent with the rise seen in the Grade 2 to Grade 7 state math scores. Without TIMSS scores before 2011 and with no way to correlate the state math test scores and TIMSS results, it is difficult to say.

To summarize: Since 2003, California has seen substantial increases in enrollment in high school science and higher level high school math. Scores on state tests at the end of those courses are also significantly higher in 2013 than in 2003. However, as measured by the 2011 TIMSS data, California's ranking in total math and numeracy was below the US average.

The 2011 TIMSS data for 9 US states allows comparison to national and international student test score in total math and numeracy for those states, However, for California and all US states, despite substantial spending for math testing since 2003, there is little to no data that would show trends over any period in achievement by state students in the arithmetic fluency needed for success in quantitative science courses.

### 3.   Requests for Test Score Data Mailed to State Department of Education

The following survey instrument was sent to public officials at state departments of education who were designated as having responsibility for responding to requests for public information, generally as stipulated in state Freedom of Information Acts (FOIA).

TO:     (State Official)

FROM:     Dr. JudithAnn R. Hartman
          Department of Chemistry
          United States Naval Academy
          572M Holloway Road
          Annapolis, MD 21402-5026

I am conducting a study of student test scores in mathematics related to preparation for chemistry.  I would like to request assistance with the following questions and information.

1.  Does (state) conduct any statewide testing in mathematics in grades 8 or higher for which nationally-normed (as opposed to criterion-referenced) results are reported?

2.  Does  (state) collect and make available any statewide testing in mathematics in grades 8 and higher that includes subtests in math (such as "procedures" or "reasoning" or "computation")?

3.  Would you be able to provide me with any document (preferably in an electronic format) that would include nationally normed student test scores in mathematics in grades 8 or higher for any period prior to 2003?  I realize this is a decade ago, but any help you could provide would be appreciated.

Thank you!

                    #  #  #  #  #

Responses to the survey were received from 29 states.  Below is summary of the responses.

Alaska:  No longitudinal computation data.
Arizona:  No computation nationally normed data.
California: No computation nationally normed data, but extensive math and science enrollment
     and state testing data since 2003 is on website.
Colorado:  No computation nationally normed data.
Delaware:  Stanford 9 and 10 data, but only limited battery "total math."
Florida:  Stanford 9 and 10 data, but only "math reasoning".
Hawaii:  No nationally normed data.
Idaho:  No nationally normed data.
Illinois:  Stanford 10 data, but only limited battery "total math."

Iowa:  ITBS data provided by Annual Condition of Education Reports to 2002, no computation since 2002.

Maryland:  No computation nationally normed data.

Massachusetts:  2011 TIMSS data provided, otherwise no computation nationally normed data.

Michigan:  No computation nationally normed data.

Missouri:  Terra Nova data, but no computation subtest report.

Montana: No computation nationally normed data.

Nevada: No computation nationally normed data.

New Mexico: Some nationally normed testing data prior to 2003, but not available.

North Dakota: CTBS/4 and CTBS/5 data available with computation subtest, 1990 to 2001.

Ohio:  No computation nationally normed data.

Pennsylvania:  No computation nationally normed data.

Rhode Island:  No computation nationally normed data.

South Dakota:  No computation nationally normed data.

Texas:  No computation nationally normed data.

Utah:  No computation nationally normed data.

Vermont:  No computation nationally normed data.

Virginia:  Stanford 9 with subtests, 1998-2002. No computation nationally normed data since 2002.

Washington State:  ITBS 1998-2003, but only total math scores reported.

Wisconsin:  No computation nationally normed data.

Wyoming  No computation nationally normed data.

# # #